\documentclass[twocolumn,showpacs,preprintnumbers,amsmath,amssymb]{revtex4-1}

\usepackage{graphicx}
\usepackage{dcolumn}
\usepackage{bm}
\usepackage{color}
\usepackage{amsmath}
\usepackage[squaren,Gray,cdot]{SIunits}

\begin{document}

\title{Photo-draining and slow capture of carriers in quantum dots probed by resonant excitation spectroscopy}

\author{Hai Son Nguyen$^1$}
\altaffiliation[Present address: ]{LPN, CNRS, 91460 Marcoussis, France.}
\author{Gregory Sallen$^1$}
\altaffiliation[Present address: ]{EPFL, CH-1015 Lausanne, Switzerland.}
\author{Marco Abbarchi$^1$}
\altaffiliation[Present address: ]{IM2NP, CNRS, AMU, 13397 Marseille, France.}
\author{Robson Ferreira$^1$}
\author{Christophe Voisin$^1$}
\author{Philippe Roussignol$^1$}
\author{Guillaume Cassabois$^{1,2,3}$}
\author{Carole Diederichs$^1$}
\email{carole.diederichs@lpa.ens.fr}

\affiliation{$^1$Laboratoire Pierre Aigrain, Ecole Normale Sup\'erieure, CNRS (UMR8551),
Universit\'{e} P. et M. Curie, Universit\'{e} D. Diderot, 24 rue Lhomond, 75231 Paris Cedex 05, France}
\affiliation{$^2$Universit\'e Montpellier 2, Laboratoire Charles Coulomb (UMR5221), 34095 Montpellier, France}
\affiliation{$^3$CNRS, Laboratoire Charles Coulomb (UMR5221), 34095 Montpellier, France}

\date{\today}

\begin{abstract}

We investigate experimentally and theoretically the resonant emission of single InAs/GaAs quantum dots in a planar microcavity. Due to the presence of at least one residual charge in the quantum dots, the resonant excitation of the neutral exciton is blocked. The influence of the residual doping on the initial quantum dots charge state is analyzed, and the resonant emission quenching is interpreted as a Coulomb blockade effect. The use of an additional non-resonant laser in a specific low power regime leads to the carrier draining in quantum dots and allows an efficient optical gating of the exciton resonant emission. A detailed population evolution model, developed to describe the carrier draining and the optical gate effect, perfectly fits the experimental results in the steady state and dynamical regimes of the optical gate with a single set of parameters. We deduce that ultra-slow Auger- and phonon-assisted capture processes govern the carrier draining in quantum dots with relaxation times in the 1 - \unit{100}{\micro\second} range. We conclude that the optical gate acts as a very sensitive probe of the quantum dots population relaxation in an unprecedented slow-capture regime.

\end{abstract}

\maketitle

\section{INTRODUCTION}

Most of the spectroscopic experiments in single quantum dots (QDs) are performed under non-resonant excitation. In such a configuration, the non-resonant laser creates carriers in the barrier or in the wetting layer. The photo-created carriers then relax to the QD excited states via \emph{capture} of the carriers from the barrier or the wetting layer to the confined QD excited states. Then, \emph{intra-dot relaxation} processes between the different QD discrete states lead to electron and hole population of the ground states from which the QD photoluminescence occurs. Theoretical works show that the capture mechanism relies either on emission of optical phonons \cite{Ferreira:99,Magnusdottir:02} with typical capture times between \unit{100}{\femto\second} and \unit{100}{\nano\second}, or on electron-electron and electron-hole Auger scattering \cite{Uskov:98,Magnusdottir:03} with typical capture times between \unit{1}{\pico\second} and \unit{1}{\micro\second}. These calculations are in perfect agreement with time-resolved experiments performed in QDs ensemble \cite{Ohnesorge:96} and in single QDs \cite{Elvira:11} where capture times of the order of \unit{100}{\pico\second} have been measured. On the other hand, irreversible Auger scattering is mainly responsible for the intra-dot relaxation process, and theoretical calculations \cite{Ferreira:99,Narvaez:06} give estimate of the corresponding relaxation times in the \unit{100}{\femto\second} - \unit{10}{\pico\second} range. In fact, differential transmission measurements \cite{Sosnowski:98}, pump-probe spectroscopy \cite{Muller:03} and time-resolved experiments \cite{Zhang:01,Siegert:05} showed that the Auger-assisted processes involved in the intra-dot carriers relaxation occur within characteristic times between \unit{1}{\pico\second} and \unit{10}{\pico\second}. In total, theoretical and experimental results show that the intra-dot relaxation times are generally much shorter (from \unit{100}{\femto\second} to \unit{10}{\pico\second}) than the characteristic times of the carriers capture from the continuum (up to \unit{1}{\micro\second}). The relaxation times to the QD ground states are thus often only attributed to the capture time. Nevertheless, intra-dot relaxation processes, although secondary in the population relaxation, play a very important role in the coherence relaxation \cite{Kammerer:02,Berthelot:06}.

The QD coherence is  another fundamental aspect that has been widely studied. At low temperature, the QD coherence is limited by spontaneous emission and QD coupling to its fluctuating environment. When spontaneous emission is the only dephasing process, the QD coherence is in the so-called radiative limit and in this ideal case, QDs appear as ideal systems to transpose the atomic physics concepts to the solid state. In this context, major experimental results have been obtained in single QDs, such as the emission of single \cite{Michler:00} and indistinguishable \cite{Santori:02} photons, the Rabi oscillations \cite{Zrenner:02} and the strong coupling regime between a single QD and an optical microcavity \cite{Reithmaier:04,Yoshie:04}. However, the radiative limit is hardly reached, showing that the intuitive artificial atom picture is strongly influenced by the QD environment.

As previously shown by C. Kammerer \emph{et al.} \cite{Kammerer:02bis}, strictly resonant excitation of single QDs is a crucial requirement in order to minimize the dephasing processes induced by the phonon and Auger-assisted carriers capture. Since the first direct measurement of the optical response of single QDs under resonant excitation performed by A. Muller \emph{et al.} \cite{Muller:07}, many results on resonant emission (RE) in single QDs were presented in the literature \cite{Melet:08,Ates:09,Vamivakas:09,Nguyen:12,Reinhard:12}. However, except for studies in the RE low power regime where the coherence is driven by the resonant excitation laser \cite{Nguyen:11,Matthiesen:12}, the radiative limit is never reached even at very low temperature. This suggests that a fluctuating electrostatic environment which influences the QD coherence still exists, even in case of resonant excitation where the laser photo-creates carriers only in the QD. Our recent work on the quenching of the QDs RE strongly corroborates this assumption \cite{Nguyen:12}. The resonant excitation of the neutral exciton in single QDs can be completely inhibited by a Coulomb blockade effect because of the presence of at least one residual charge in the QD. However, this issue is overcome with the use of an additional weak non-resonant laser which neutralizes the QD and optically gates its RE \cite{Nguyen:12}, as also shown in other recent experimental studies \cite{Muller,Reinhard:12,Volz:12}. The use of an additional weak non-resonant laser also appears to be interesting to probe the local environment fluctuations at the single-charge level in single QD \cite{Houel:12} and to minimize this environment fluctuations to improve the emitted photons indistinguishability \cite{Gazzano:13}.

The present paper is devoted to a comprehensive experimental and theoretical study of this optical gating effect in single QDs under resonant excitation. The influence of the residual doping on the initial QD charge state, and consequently on the QD RE quenching, is analyzed. Moreover, a detailed population evolution model is developed to describe the carrier draining by the weak optical gate in the QD. This model which perfectly fits, with a single set of parameters, the experimental results in the steady state and dynamical regimes, shows that the carrier draining in the QD is governed by peculiar Auger- and phonon-assisted capture processes. In fact, beyond the explanation of the optical gate effect, the original weak optical gate configuration allows us to study an unprecedented regime where the carriers capture involved in the QD draining is governed by ultra-slow processes with time constants of the order of 1 to \unit{100}{\micro\second}. Therefore, the optical gate appears to be a very sensitive probe of the residual doping of the sample and of the QD population in a regime of a very weak non-resonant excitation.

The paper is organized as follows: In section~\ref{setup}, we give some details on the sample and the experimental configuration. Section~\ref{gate phenomenology} is devoted to the resonant excitation results on a single QD, showing a striking quenching of the neutral exciton RE which appears to be a general behavior. When such a situation is observed, the use of an additional non-resonant laser that acts as an optical gate allows a complete retrieval of the neutral exciton RE. In parallel in appendix, we develop a self-consistent calculation of the influence of the residual doping on the QD charge state in the absence of optical excitation. The calculations show that the RE quenching can be interpreted as a Coulomb blockade effect where at least one residual charge in the QD blocks the resonant excitation of the neutral exciton. The optical gate phenomenology is then presented in an experimental study of the RE intensity and energy shift as a function of the optical gate power. Section~\ref{theory} presents a population evolution model where the different carriers capture and escape processes induced by the optical gate are considered. Section~\ref{theory vs experiment} is finally devoted to a confrontation of our theoretical model to experimental data on the optically-gated RE obtained either in the steady state or dynamical regimes. Distinct experiments allow to obtain a highly constrained set of parameters used in our model. Finally, as an illustration of the robustness of our approach, we reproduce quantitatively measurements of the emitted photon statistics. This statistics shows the usual antibunching at zero delay, but a strong photons bunching at larger delays. This bunching is essentially due to fluctuations of the QD between a neutral and a charged state.

\section{\label{setup}SAMPLE DESCRIPTION AND EXPERIMENTAL SETUP}

Our sample is made of a single layer of self-assembled InAs/GaAs QDs inserted in an AlAs/AlGaAs planar microcavity. The QDs density varies from $10^7$ to $10^{10}$~QDs/{\centi\meter\squared} with an energy distribution of the emission from \unit{1.240}{\electronvolt} to \unit{1.305}{\electronvolt} with a maximum at \unit{1.270}{\electronvolt}. In the experiments, we focus on the low density region of the sample where the QDs density is about $10^{8}$~QDs/{\centi\meter\squared}. The planar microcavity is a Fabry-Perot cavity made of two Bragg mirrors which are composed of alternating AlAs and AlGaAs layers of identical optical thicknesses $\lambda_0/4$ ($\lambda_0=1$~{\micro\meter}). The cavity spacer is a $\lambda_0$-GaAs layer and the top and bottom Bragg mirrors are respectively composed of 11 and 24 AlAs/AlGaAs pairs, resulting in a quality factor of 2500. The Fabry-Perot cavity mode is then centered at \unit{1.270}{\electronvolt} with a full width at half maximum of \unit{0.55}{\milli\electronvolt}. The cavity asymmetry facilitates the extraction of the QD emission in the cavity mode in our conventional confocal micro-photoluminescence setup which is used in a reflection configuration for the sample characterization under non-resonant excitation. The microscope objective used in this setup has a numerical aperture NA=0.35 leading to a spatial resolution of \unit{2}{\micro\metre} required for experimental studies on single QDs. The resonant excitation of single QDs is performed at \unit{7}{\kelvin} by spatially decoupling the excitation from the detection paths in order to get rid of the elastic laser scattering \cite{Nguyen:11,Nguyen:12}. The excitation of the QDs is done along the lateral facet of the sample by using a lensed fiber inserted in the cryostat while the detection is performed in the vertical direction with the confocal setup. The sample is fixed on a piezo-electric stage in order to optimize its position with respect to the lensed fiber. Once coupled into the sample, the laser is confined in the microcavity and propagates. The planar microcavity thus acts as a two-dimensional waveguide for the excitation laser and the main contributions to the laser scattering arises from the edge of the sample or the defects in the Bragg mirrors. Since the results presented in this paper concern the RE of a QD located far away from the edge of the sample, approximately at \unit{1}{\milli\meter}, the detection of the QD emission is not disturbed by the laser scattering on the edge of the sample. In fact, as we will see in the experimental results, the laser scattering is much weaker, even negligible, than the QD signal emission.
The resonant excitation laser is a cw external cavity laser diode tunable from 1.25 to \unit{1.29}{\electronvolt} (i.e. from 960 to \unit{990}{\nano\meter}) with a spectral resolution of \unit{0.5}{\micro\electronvolt} and with a spectral linewidth of \unit{1.25}{\nano\electronvolt} (i.e. \unit{300}{\kilo\hertz}).

\section{\label{gate phenomenology}PHENOMENOLOGY OF THE OPTICAL GATE EFFECT}
\subsection{Quantum dot emission under non resonant excitation}

\begin{figure}[htb]
\begin{center}
\includegraphics[width=8.6cm]{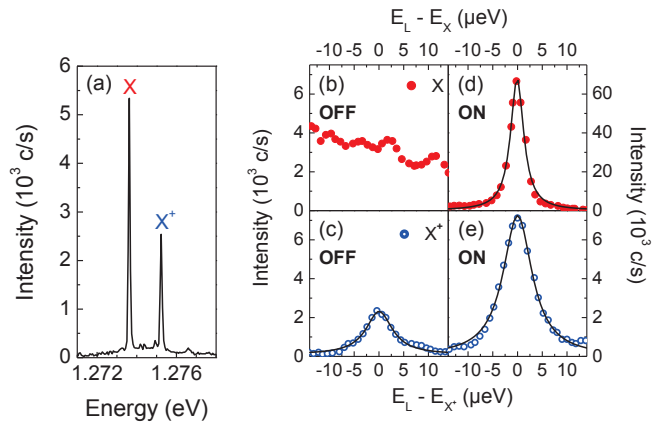}
\end{center}
\caption{\label{Figure1}{(a) Typical spectrum of the non-resonantly excited photoluminescence of a single quantum dot, showing the neutral exciton $X$ and the positive trion $X^+$. (b,c,d,e) Resonant emission spectra of $X$ ($\bullet$) and $X^+$ ($\circ$) as a function of the detuning between the laser energy $E_L$ and the exciton (trion) energy $E_X$ ($E_{X^+}$) when the optical gate is switched off (b,c) and on (d,e).}}
\end{figure}

Under non-resonant excitation in the GaAs barrier (with a He-Ne laser), the photoluminescence spectra exhibit two distinct lines typically spaced by \unit{1.5}{\milli\electronvolt}. Figure~\ref{Figure1}(a) shows a typical non-resonant photoluminescence spectrum for a \unit{10}{\micro\watt} excitation (the excitation power is measured before the microscope objective). Power-dependent studies and photon-correlation measurements \cite{Nguyen:12} were performed to identify the two lines as a neutral exciton $X$, at the energy $E_X=1.2736$~{\electronvolt}, and a charged exciton at the energy $E_{X^+}=1.2752$~{\electronvolt}. Moreover, the photoluminescence of the GaAs barrier (not shown here) shows an emission line at \unit{1.495}{\electronvolt} resulting from the radiative recombination of the free electrons in the barrier with the holes bound to the carbon acceptors that are the unintentional impurities which appear during the sample growth \cite{Hayne:04}. These impurities are notably responsible for the presence of residual holes in the sample. Thus we consider that the charged exciton is a positive trion $X^+$, which is further supported by the positive detuning of the charged exciton state with respect to the neutral one \cite{Regelman:01,Ediger:05}. The observation of the positive trion in the non-resonant photoluminescence spectra shows that the QDs in the studied sample may efficiently capture residual holes that exist in the QDs environment. Nevertheless, under these excitation conditions, the simultaneous observation of the two lines indicates that the QD is either empty or populated by at least one hole. The situation is completely different under strictly resonant excitation.

\subsection{Quenching of the neutral exciton resonant emission}

In fact, we have recently reported on the quenching of the resonant emission of single QDs at the energy of the neutral exciton \cite{Nguyen:12}. As shown in figure~\ref{Figure1}(b) where the RE is plotted as a function of the laser detuning with respect to the neutral exciton energy, no resonance is observed at zero detuning whereas the corresponding experiment for the positive trion shows clearly a signal enhancement when the laser energy perfectly matches its energy (Fig.~\ref{Figure1}(c)). A statistical study showed that for more than 90\% of the QDs, the RE signal arises only from the positive trion. This is due to the presence of at least one residual hole in the QDs. These residual charges induce a photon absorption at the energy of charged excitonic complexes and not at the neutral exciton one for which the QD must be empty. This phenomenon is similar to the Coulomb blockade effect where the creation of a neutral exciton is blocked by the residual hole in the QD \cite{Nguyen:12}. This difficulty can be circumvented by the use of a very weak additional non-resonant laser (optical gate).

\subsection{Control of the resonant emission by an optical gate}

Starting from the experimental configuration used for the QDs resonant spectroscopy (see section~\ref{setup}), an additional laser (He-Ne laser) is injected perpendicularly to the sample surface via the confocal microscopy setup. This additional laser acts as an optical gate for the QD RE \cite{Nguyen:12,Muller,Reinhard:12,Volz:12}. Because of the very low gate power ($P_{\text{gate}}\sim3$~{\nano\watt} in Fig.~\ref{Figure1}(d,e)), the non-resonant laser does not populate significantly the QD. Power-dependent studies (presented in section~\ref{RE vs gate power}, Fig.~\ref{Figure2}(a)) show indeed that the photoluminescence induced by the HeNe laser cannot be distinguished from the noise background for a typical excitation power smaller than \unit{0.1}{\micro\watt}. Optical gating of the QD RE of the neutral exciton and the positive trion are presented in figures~\ref{Figure1}(d,e). A strong RE signal is now detected for the neutral exciton with an enhancement factor of about 30. These data show that an ultra-weak non-resonant laser tends to neutralize the QD and then allows the resonant excitation of the neutral exciton. This optical gate also acts on the trion with an enhancement factor of 3. As a consequence, we assume that the QD contains possibly more than one hole when the gate is off. This assumption is supported by the self-consistent calculation of the QD population presented in appendix.

\subsection{\label{RE vs gate power}Resonant emission intensity versus the optical gate power}

\begin{figure}[htb]
\begin{center}
\includegraphics[width=8.6cm]{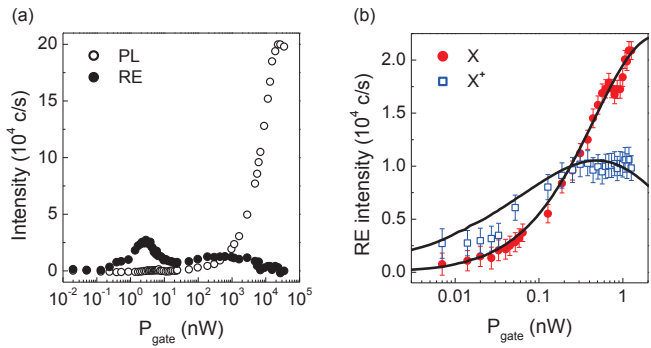}
\end{center}
\caption{\label{Figure2}{(a) Resonant emission (RE) and non-resonant photoluminescence (PL) intensities of the neutral exciton ($X$) as a function of the optical gate power $P_{\text{gate}}$. (b) Resonant emission intensity of the neutral exciton ($X$) and the positive trion ($X^+$) as a function of the optical gate power $P_{\text{gate}}$. Only the region where the non-resonant photoluminescence is completely negligible is displayed. The experimental data are fitted by the population evolution model described in section~\ref{theory} (solid lines).}}
\end{figure}

Figure~\ref{Figure2}(a) presents the intensity of the optically-gated RE and the non-resonant photoluminescence of the exciton as a function of the optical gate power over seven orders of magnitude. The photoluminescence signal, which starts to be significant at $P_{\text{gate}}\sim0.1$~{\micro\watt}, shows a standard power dependence with a saturation at $P_{\text{gate}}\sim30$~{\micro\watt}. Concerning the optically-gated RE, measured for a constant resonant excitation power of \unit{5}{\micro\watt} (the saturation regime appears for a resonant excitation power of \unit{16.5}{\micro\watt} \cite{Nguyen:12}), three remarkable features can be distinguished. First, the RE appears at $P_{\text{gate}}\sim0.1$~{\nano\watt} and increases to a maximum value $I_{max}\sim3.10^4$~counts/s for $P_{\text{gate}}\sim3$~{\nano\watt} while in this region the non-resonant photoluminescence remains below the detection threshold. Once the maximum signal is reached, the optically-gated RE rapidly decreases for gate powers ranging between \unit{3}{\nano\watt} and \unit{30}{\nano\watt}. At this latter excitation power, the intensity of non-resonant photoluminescence starts to be detectable. From \unit{0.1}{\micro\watt}, the photoluminescence excited by the optical gate becomes significant while the RE decreases and becomes negligible at $P_{\text{gate}}\sim30$~{\micro\watt} where the QD is saturated.

Figure~\ref{Figure2}(b) shows the RE intensities of the exciton and the positive trion as a function of the optical gate power in the restricted range \unit{0.007-2}{\nano\watt}. In this range the non-resonant photoluminescence excited by the gate is completely negligible. For gate powers smaller than \unit{0.5}{\nano\watt}, the trion RE increases with gate power indicating that the residual QD occupation evolves from two to one residual holes. For gate powers larger than \unit{0.5}{\nano\watt} the trion RE saturates before a preliminary decrease, the QD becoming empty.

We have seen that the optical gate substantially modifies the QD population, we will now consider the consequences related to the QD electrostatic environment.

\subsection{\label{shift}Energy of the resonant emission versus the optical gate power}

\begin{figure}[htb]
\begin{center}
\includegraphics[width=8.6cm]{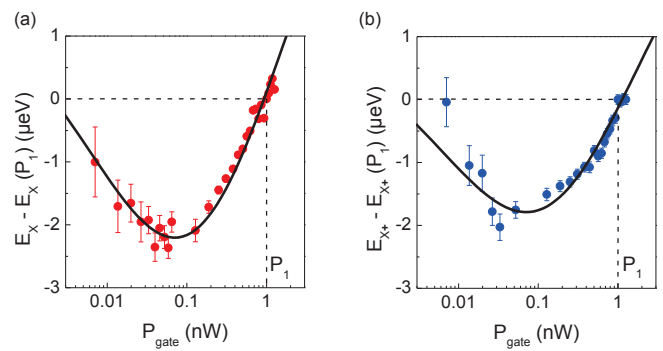}
\end{center}
\caption{\label{Figure3}{Energy of the neutral exciton (a) and positive trion (b) resonant emission as a function of the optical gate power $P_{\text{gate}}$. The experimental data are obtained from the same set of measurements as figure~\ref{Figure2}(b) and are fitted by the model described in section~\ref{theory} (solid lines).}}
\end{figure}

Figure~\ref{Figure3} displays the RE energies of the neutral exciton, $E_X$ (Fig.~\ref{Figure3}(a)), and the positive trion, $E_{X^+}$ (Fig.~\ref{Figure3}(b)), as a function of the optical gate power $P_{\text{gate}}$. The solid lines are obtained from the theoretical model described in details in the next section. A red shift of few {\micro\electronvolt}s is first observed in the very weak gate power regime, followed by a comparable blue shift when the gate power increases. The transition between the red and  blue shift occurs at a typical power of $P_{\text{gate}}\sim0.05$~{\nano\watt}. This optical gate power is much smaller than the gate powers corresponding to the maximum RE intensities of the exciton ($P_{\text{gate}}\sim3$~{\nano\watt}) and the trion ($P_{\text{gate}}\sim0.5$~{\nano\watt}). In fact, the energy shifts of $X$ and $X^+$ are governed, through the quantum confined Stark effect, by the charge state of the QD environment, while the RE intensity only depends on the charge state of the QD itself. This point will be quantitatively addressed in the next section.

\section{\label{theory}THEORETICAL MODEL OF THE OPTICAL GATE EFFECT}

In the following, we propose a simple theoretical model which describes the processes involved in the quenching of the neutral exciton RE  and the operation of the optical gate.

\subsection{\label{pop without gate}Population evolution without the optical gate}

\begin{figure}[htb]
\begin{center}
\includegraphics[width=8.6cm]{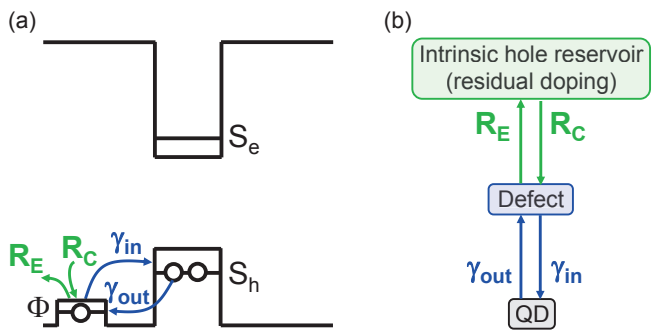}
\end{center}
\caption{\label{Figure4}{(a) Scheme of a QD with a fundamental hole state $S_h$ quasi-resonant with a slightly confined interfacial defect state $\Phi$. The QD is filled with two holes and the absorption of a photon at the neutral exciton energy is impossible. (b) The capture and escape processes are schematized with the corresponding rates in a diagram showing the intrinsic hole reservoir induced by the residual doping of the sample (see appendix), the defect and the QD.}}
\end{figure}

As explained in appendix, the residual doping of the sample is at the origin of a hole reservoir which will notably influence the QD emission. As we do not have a precise knowledge of the QD electrostatic environment, we simplify the general description of the influence of the residual doping on the QD charge state by considering an effective model where the distribution of holes at the vicinity of the QD is schematized by a single trap such as an interface defect of the wetting layer (see Fig.~\ref{Figure4}). Residual holes introduced by the non-intentional doping of the sample will preferably be trapped (de-trapped) in (from) the defect, and the corresponding capture (escape) rate is denoted $R_C$ ($R_E$). For the sake of simplicity, each QD is supposed to be coupled to only one defect (the phenomenology is the same in case of more defects). This defect is located in the vicinity of the QDs and a tunneling channel between the defect state $\Phi$ and the QD fundamental hole state $S_h$ opens if the defect is close enough to the QD (the distance between the QD and the interface defect is discussed in section \ref{theory_shift}). The hole tunneling rate from the defect to the QD is denoted $\gamma_{in}$, and $\gamma_{out}$ for the inverse process rate. Because the confinement of an interface defect is generally small, the $\Phi$ state is close to the valence band continuum and we assume that $\gamma_{in}\gg\gamma_{out}$. Finally, due to the spin of the carriers, the defect state $\Phi$ and the QD fundamental electron and heavy hole states, $S_e$ and $S_h$, are twice degenerated and can then be populated by up to two carriers. From our experimental observations and from the calculation developed in appendix, this assumption seems reasonable since, without any light illumination, the $S_h$ state of the QD is populated in average by two holes at most.

Within these hypothesis, the evolutions of the average number of holes in the defect, $n_p$, and the average number of holes (electrons) in the QD, $n_h$ ($n_e$), are described by the rate equations:
\begin{subequations}
\begin{eqnarray}
\frac{dn_p}{dt}&=&R_C(2-n_p)-R_En_p-\gamma_{in}(2-n_h)n_p\label{trou_piege_gate_off}\\
&+&\gamma_{out}(2-n_p)n_h\nonumber\\
\frac{dn_h}{dt}&=&\gamma_{in}(2-n_h)n_p-\gamma_{out}(2-n_p)n_h\label{trou_boite_gate_off}\\
\frac{dn_e}{dt}&=&0
\end{eqnarray}
\end{subequations}

Considering $\gamma_{in}\gg\gamma_{out}$, the populations in the steady state are then given by $n^{(st)}_p=2R_C/(R_C+R_E)$, $n^{(st)}_h\approx2$ and $n^{(st)}_e=n_e=0$, where two holes populate the $S_h$ state with the corresponding QD fundamental state $|2_h;0_e\rangle$. Therefore, the absorption of a photon at the energy of the neutral exciton, for which the ground state $|g\rangle$ is $|0_h;0_e\rangle$, is impossible and an inhibition of the neutral exciton RE is observed. A priori, such configuration neither allows the absorption of a photon at the energy of the positive trion, which is associated to the fundamental state $|1_h;0_e\rangle$, and the trion RE should not be observed contrarily to what is shown in Fig.~\ref{Figure1}(c). However, $n^{(st)}_p$ and $n^{(st)}_h$ are only the average numbers of carriers in the steady state, and, in the dynamical regime, fluctuations of $n_p$ and $n_h$ associated to a hole exchange between the QD and the defect exist. Therefore, even though the QD ground steady state is $|2_h;0_e\rangle$, the $|1_h;0_e\rangle$ state is sometimes reachable when a hole escapes from the QD to the defect. The resonant excitation of $X^+$ then becomes possible with nevertheless a low efficiency.

This elementary model explains the phenomenology of the quenching of the neutral exciton RE while the trion RE can be observed. In next section, we improve this model in order to obtain a quantitative description of the optical gate effect.

\subsection{\label{pop with gate}Population evolution with the optical gate}

\begin{figure}[htb]
\begin{center}
\includegraphics[width=8.6cm]{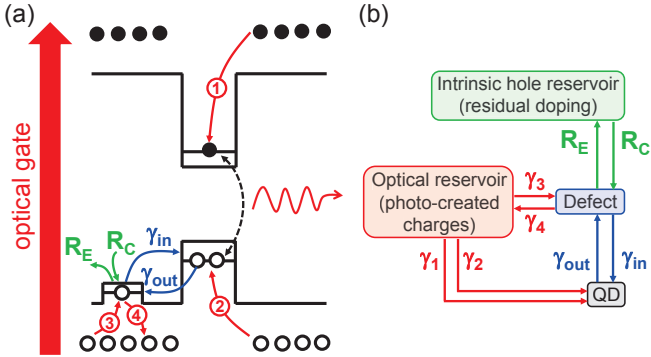}
\end{center}
\caption{\label{Figure5}{(a) Sketch of the optical gate effect: carriers are created in the GaAs barrier and captured by the QD or the defect. (b) The capture and escape processes induced by the optical gate are schematized in a diagram showing the optical reservoir formed by the charges photo-created in the GaAs barrier by the optical gate.}}
\end{figure}

The optical gate populates the GaAs barrier as described in figure~\ref{Figure5} in the absence of the resonant excitation laser. The evolution of the average number of holes in the defect, $n_p$, in the QD, $n_h$, and the average number of electrons in the QD, $n_e$, is now described by the following rate equations:
\begin{subequations}\label{evolution_general}
\begin{eqnarray}
\frac{dn_p}{dt}&=&R_C(2-n_p)-R_En_p-\gamma_{in}(2-n_h)n_p\\
&+&\gamma_{out}(2-n_p)n_h+\gamma_3(2-n_p)-\gamma_4n_p\nonumber\\
\frac{dn_h}{dt}&=&\gamma_{in}(2-n_h)n_p-\gamma_{out}(2-n_p)n_h\\
&+&\gamma_2(2-n_h)-\gamma n_h n_e\nonumber\\
\frac{dn_e}{dt}&=&\gamma_1(2-n_e)-\gamma n_h n_e
\end{eqnarray}
\end{subequations}
where $\gamma_{1}$ stands for the electron capture in the QD and $\gamma_{2}$ the hole capture rate in the QD, both from the barrier. The rates  $\gamma_{3}$ and $\gamma_{4}$ correspond to the hole capture and escape for the defect, respectively. Since we study QDs in the strong confinement regime, the carriers trapped in the QD form electron-hole pairs in Coulomb interaction characterized by a relatively small binding energy compared to the kinetic energy resulting from the geometric confinement. In this context and in case of a low gate power for which the QD is hardly populated by more than one electron-hole pair in the fundamental state, the radiative recombination of an electron-hole pair in the QD at the neutral exciton energy is proportional to $n_e n_h$ when spin effects are not considered. The related radiative recombination rate $\gamma$ corresponds to an excitonic radiative lifetime $T_1=330$~{\pico\second}, which has been measured by time-resolved photoluminescence experiments in the same experimental configuration \cite{Nguyen:12}.

In the explored power range, we assume the following hierarchy  between the rates in the set of equations~(\ref{evolution_general}):
\begin{equation}\label{hierarchy}
    R_C,R_E,\gamma, \gamma_1, \gamma_2, \gamma_3, \gamma_4\gg\gamma_{in}\gg\gamma_{out}
\end{equation}
This assumption will be justified in the following (see section~\ref{on to off}). Within this hypothesis, the variations of the QD charge state is essentially governed by the two capture processes (1), (2), and its radiative recombination. Concerning the defect charge state, it is governed by the two capture processes from the barrier and from the hole reservoir ($\gamma_3$ and $R_C$), and the corresponding two escape processes ($\gamma_4$ and $R_E$).

Consequently, in the absence of the resonant excitation laser, the evolution of the charge states of the QD and the defect becomes independent when the optical gate is switched on, and the set of equations~(\ref{evolution_general}) is reduced to:
\begin{subequations}
\begin{eqnarray}
\frac{dn_p}{dt}&=&R_C(2-n_p)-R_En_p+\gamma_3(2-n_p)-\gamma_4n_p\label{trou_piege}\\
\frac{dn_h}{dt}&=&\gamma_2(2-n_h)-\gamma n_h n_e\label{trou}\\
\frac{dn_e}{dt}&=&\gamma_1(2-n_e)-\gamma n_h n_e\label{electron}
\end{eqnarray}
\end{subequations}

\subsection{Excitonic complexes population evolution}

In fact, under resonant excitation, the equations (\ref{trou}) and (\ref{electron}) are not satisfied since they do not take into account the electron and hole populations that are created by the resonant excitation laser. In this context, the QD charge state drastically modifies the absorption of the resonant excitation laser. We stress that the model is developed for resonant excitation powers where the resonance fluorescence contribution in the RE overwhelms the resonant Rayleigh scattering \cite{Nguyen:11}. We thus focus on the populations and do not take into account any coherent effect. Moreover, the model describes the linear regime before the saturation of the QD by the resonant laser, which corresponds to the actual experimental configuration. Therefore, this model is only valid for intermediate powers of the resonant excitation laser, where stimulated emission processes are not considered, corresponding to its weak coupling regime with the QD (i.e., the strong coupling regime between the resonant laser and the QD, characterized by the so-called Mollow triplet, is not considered). Finally, this model is suitable for a non-resonant optical gate where the photo-created holes and electrons are not correlated.

\begin{figure}[htb]
\begin{center}
\includegraphics[width=5cm]{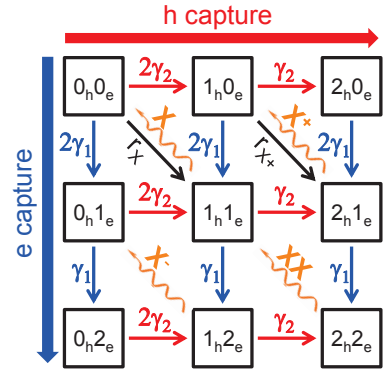}
\end{center}
\caption{\label{Figure6}{Diagram of the population evolution model.}}
\end{figure}

In order to consider the different carrier populations in the QD, we adopt a population evolution model similar to the one proposed by J. Gomis-Bresco \emph{et al.} \cite{Gomis:11}, which adapts for excitons the random population model of M. Grundmann \emph{et al.} \cite{Grundmann:97}. Within this approach, all the processes related to the charge capture, the radiative recombination, and the resonant pumping are taken into account by considering all the possible charge configurations of the QD. We restrict ourselves to the neutral exciton $X$, the positive and negative trions $X^{\pm}$ and the biexciton $XX$. These processes are schematized in figure~\ref{Figure6}, where the radiative recombination rates of these states are respectively denoted $\gamma_X$, $\gamma_{X^{\pm}}$ and $\gamma_{XX}$ with $\gamma_X=\gamma_{X^{\pm}}\equiv\gamma$ and $\gamma_{XX}=2\gamma_X$ \cite{Gomis:11}. The resonant pumping rates at the energy of the neutral exciton $X$ and the positive trion $X^+$ are denoted $r_X$ and $r_{X^+}$ respectively, and depend on the power and the wavelength of the resonant excitation laser. Depending on the experimental configuration (i.e., resonant excitation of $X$ or $X^+$), either $r_X$ or $r_{X^+}$ are set to zero in order to consider the resonant excitation of only one of these states. The probability ${\cal P}_{i_{h};j_{e}}$ ($i,j=0,1$ and 2) to obtain $i_h$ holes and $j_e$ electrons in the QD is calculated by replacing the equations (\ref{trou}) and (\ref{electron}) by the following linear differential equations system:
\begin{equation}
\frac{d\cal P}{dt}=\cal T.P \label{mpa}
\end{equation}
where
\begin{eqnarray}
{\cal P}=\left(
    \begin{array}{c}
    {\cal P}_{0_h;0_e}\\
    {\cal P}_{0_h;1_e}\\
    {\cal P}_{0_h;2_e}\\
    {\cal P}_{1_h;1_e}\\
    {\cal P}_{1_h;2_e}\\
    {\cal P}_{2_h;1_e}\\
    {\cal P}_{2_h;2_e}\\
    {\cal P}_{2_h;0_e}\\
    {\cal P}_{1_h;0_e}
    \end{array}
\right)
\text {with $\sum_{(m,n)}{\cal P}_{i_h;j_e}=1$}
\end{eqnarray}
and
\begin{widetext}
\begin{equation}
{\cal T}=\left( \begin{matrix}
	-2(\gamma_1+\gamma_2)-r_X & 0 & 0 & \gamma & 0 & 0 & 0 & 0 & 0\\
    2\gamma_1 & -(\gamma_1+2\gamma_2) & 0 & 0 & \gamma & 0 & 0 & 0 & 0\\
    0 & \gamma_1 & -2\gamma_2 & 0 & 0 & 0 & 0 & 0 & 0\\
    r_X & 2\gamma_2 & 0 & -(\gamma+\gamma_1+\gamma_2) & 0 & 0 & \gamma & 0 & 2\gamma_1\\
    0 & 0 & 2\gamma_2 & \gamma_1 & -(\gamma_2+\gamma) & 0 & 0 & 0 & 0\\
    0 & 0 & 0 & \gamma_2 & 0 & -(\gamma+\gamma_1) & 0 & 2\gamma_1 & r_{X^+}\\
    0 & 0 & 0 & 0 & \gamma_2 & \gamma_1 & -\gamma & 0 & 0\\
    0 & 0 & 0 & 0 & 0 & 0 & 0 & -2\gamma_1 & \gamma_2\\
    2\gamma_2 & 0 & 0 & 0 & 0 & \gamma & 0 & 0 & -(2\gamma_1+\gamma_2)-r_{X^+}
\end{matrix}\right)
\end{equation}
\end{widetext}
The evolution of any charge state can be calculated in order to explain the RE signal observed at low or high power of the optical gate. Nevertheless, the validity of this model is limited to the case where the tunnel effect between the QD and the defect is negligible, meaning that the ultra-low power regime, where $\gamma_{in}$ and $\gamma_{out}$ are comparable to $\gamma_1$ and $\gamma_2$, cannot be explained here.

\section{\label{theory vs experiment}THEORY VERSUS EXPERIMENTAL RESULTS}

\subsection{\label{theory RE vs gate power}Photoluminescence and resonant emission dependence on the optical gate power}

In our model, the intensity of the photoluminescence non-resonantly excited by the optical gate, $I_X^{PL}$, is proportional to the population of the QD state $|1_h;1_e\rangle$ with resonant pumping rates equal to zero, leading to:
\begin{equation}\label{PL_X}
    I_X^{PL}\propto\left\langle \mathcal{P}_{1_h;1_e}\right\rangle_{(st)}\Big\lvert_{r_{X}=r_{X^+}=0}
\end{equation}
The optically-gated RE of the exciton, $I_X^{RE}$, is proportional to the difference of populations in the QD state $|1_h;1_e\rangle$ with and without a resonant pumping $r_X=r$, such as:
\begin{equation}\label{RE_X}
    I_X^{RE}\propto\left\langle \mathcal{P}_{1_h;1_e}\right\rangle_{(st)}\Big\lvert_{r_X=r,\,r_{X^+}=0} -\left\langle \mathcal{P}_{1_h;1_e}\right\rangle_{(st)}\Big\lvert_{r_X=r_{X^+}=0}
\end{equation}
Likewise, the RE of the positive trion, $I_{X^+}^{RE}$, is defined from the populations difference in the QD state $|2_h;1_e\rangle$ with and without a resonant pumping $r_{X^+}=r$, such as:
\begin{equation}\label{RE_X+}
    I_{X^+}^{RE}\propto\left\langle \mathcal{P}_{2_h;1_e}\right\rangle_{(st)}\Big\lvert_{r_X=0,\,r_{X^+}=r} -\left\langle \mathcal{P}_{2_h;1_e}\right\rangle_{(st)}\Big\lvert_{r_X=r_{X^+}=0}
\end{equation}

\begin{figure}[htb]
\begin{center}
\includegraphics[width=8.6cm]{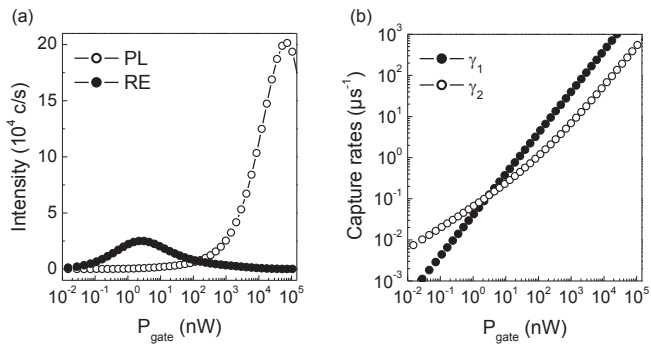}
\end{center}
\caption{\label{Figure7}{(a) Results of the simulation showing the resonant emission (RE) and the non-resonant photoluminescence (PL) intensities of the neutral exciton as a function of the optical gate power $P_{\text{gate}}$. (b) Electron and hole capture rates, $\gamma_1$ and $\gamma_2$, as a function of the optical gate power $P_{\text{gate}}$.}}
\end{figure}

The two first equations (\ref{PL_X}) and (\ref{RE_X}) are used to reproduce the experimental results in figures~\ref{Figure2}(a,b) and equation~(\ref{RE_X+}) is used to reproduce the RE arising from the trion in figure~\ref{Figure2}(b). The results of the simulation, presented in figures~\ref{Figure2}(b) and \ref{Figure7}(a) are obtained with the four following parameters:
\begin{eqnarray}
\left\lbrace
    \begin{array}{l}
        r=15~{\micro\reciprocal\second}\\
        \gamma=1/T_1=3030~{\micro\reciprocal\second}\\
        \gamma_{1({\micro\reciprocal\second})}=0.04P_{\text{gate}({\nano\watt})}\\
        \gamma_{2({\micro\reciprocal\second})}=0.06\sqrt{P_{\text{gate}({\nano\watt})}}+0.005P_{\text{gate}({\nano\watt})}\label{gamma12}
    \end{array}
\right.
\end{eqnarray}
In fact, the experimental results depicted in figures~\ref{Figure2}(a,b) only allow to determine the ratio $\gamma_1/\gamma_2$. The absolute values of these parameters are deduced from additional photon-correlation measurements that are presented in section \ref{g2}. By using these parameters, the PL and RE intensity dependencies on the gate power are fairly reproduced both for the exciton $X$ or the trion $X^{+}$. We stress here that the values of the above four parameters are used in all the simulations presented in the paper.

Let us now discuss the power dependence introduced for $\gamma_{1}$ and $\gamma_{2}$. Considering the evolution of the barrier electron (hole) population $N_e$ ($N_h$) under a non-resonant excitation \cite{Pelant:12} and the fact that the photoluminescence of the GaAs barrier is not influenced by the QDs population \cite{Cassabois:02}, the populations $N_e$ and $N_h$ created by the optical gate are governed by bimolecular interband radiative recombination \cite{Berthelot:06,Pelant:12} and are proportional to $\sqrt{P_{\text{gate}}}$. The carriers can be then captured in the QD and the defect in its vicinity, either by the emission of optical phonons, or by Auger processes. The capture of a carrier in the QD by the emission of an optical phonon is a single charge process and is therefore proportional to the number of electrons or holes in the continuum \cite{Ferreira:99,Raymond:00,Magnusdottir:02}, and in other words to $\sqrt{P_{\text{gate}}}$. Contrarily to this latter process, the Auger-assisted capture of a carrier involves two charges, where the incident charge in the continuum interacts with a target charge either in the continuum (Auger effect of type (I)), or in the QD (Auger effect of type (II)). Here, we consider Auger processes of type (I) because the type (II) processes are only efficient for a small range of QDs sizes \cite{Magnusdottir:03,Uskov:98}. The corresponding capture rate is then proportional to $P_{\text{gate}}$. Symmetrically to these capture processes, holes can escape from the defect by absorption of a phonon or by an Auger effect. De-trapping from the QD  hardly occurs considering the high values of the hole and electron confinement energies (about 30 and \unit{40}{\milli\electronvolt}, respectively). Even though we do not have a precise knowledge of the electrons and holes confinements in our sample, we assume that the carriers confinement is similar to the one in samples that were fabricated in the same facilities and which show a photoluminescence signal at the same excitonic energy (i.e. \unit{1.27}{\electronvolt}) \cite{Medeiros:95}. In this type of samples, the confinement energy of the first electron (hole) state is larger (smaller) than the energy of an optical phonon so that the holes capture is, contrarily to the electrons, efficiently assisted by the emission of optical phonons. Moreover, regarding the linear power dependence of $\gamma_1$ and $\gamma_2$ at high gate powers when the non-resonant photoluminescence is observed, we conclude that the capture processes are completely governed by Auger processes. This observation is also in good agreement with the experimental work of B. Ohnesorge \emph{et al.}~\cite{Ohnesorge:96} showing that the capture processes are mainly phonon-assisted at low excitation density and Auger-assisted at high excitation density.

In order to compare the capture rates used in our fits to the ones given in literature \cite{Uskov:98,Magnusdottir:02,Magnusdottir:03}, the charge density has to be considered instead of the optical excitation power. Considering the number of electron-hole pairs created by the optical gate in the GaAs barrier $N=\sqrt{\frac{C}{\gamma_{GaAs}}}\sqrt{P_{\text{gate}}}$ \cite{Berthelot:06,Pelant:12}, with $C$ the photo-generation coefficient and $\gamma_{GaAs}$ the radiative recombination rate of the electron-hole pairs in the GaAs barrier which are respectively about $10^{10}$~{\nano\reciprocal\watt\reciprocal\second} (for a HeNe laser at \unit{1.96}{\electronvolt}) and \unit{1}{\nano\reciprocal\second} \cite{Gobel:85}, the density of charges photo-created in the GaAs barrier is given by $n_{(\meter\rpsquared)}\approx10^{12}\sqrt{P_{\text{gate}({\nano\watt})}}$ (for a laser spot diameter of \unit{2}{\micro\meter}). Therefore, the power laws of $\gamma_1$ and $\gamma_2$ can be rewritten as:
\begin{eqnarray}
\left\lbrace
    \begin{array}{l}
        \gamma_1=C_e^{Auger}n^2\\
        \gamma_2=C_h^{Auger}n^2+A_h^{LO}n
    \end{array}
\right.
\end{eqnarray}
where $C_e^{Auger}=4\times10^{-20}$~\power{\meter}{4}{\per\second} and $C_h^{Auger}=5\times10^{-21}$~\power{\meter}{4}{\per\second} are respectively the coefficients for the electron and the hole capture by Auger effect, and $A_h^{LO}=6\times10^{-8}$~{\meter\squared\per\second} is the coefficient for the hole capture by emission of optical phonons. The estimates of the Auger coefficients are in agreement with the calculations of A. V. Uskov \emph{et al.} \cite{Uskov:98} which predict an Auger coefficient of $2\times10^{-20}$~\power{\meter}{4}{\per\second}, as well as with the calculations of I. Magnusdottir \emph{et al.} \cite{Magnusdottir:03} which give coefficients of $10^{-22}-10^{-20}$~\power{\meter}{4}{\per\second} for InAs/GaAs QDs with radius of \unit{5-15}{\nano\meter}. Concerning the phonon assisted capture, our estimated value of $A_h^{LO}$ is equivalent to the coefficient $\widetilde{A}_h^{LO}=e_{GaAs}A_h^{LO}=1.8~10^{-5}$~\power{\centi\meter}{3}{\per\second} where the substrate thickness of the sample, $e_{GaAs}\approx300$~{\micro\meter}, is taken into account. This value is also in perfect agreement with the calculations of I. Magnusdottir \emph{et al.} \cite{Magnusdottir:02} where $\widetilde{A}_h^{LO}=3.8~10^{-5}$~\power{\centi\meter}{3}{\per\second}.

Figure~\ref{Figure7}(b) displays the dependence on the optical gate power $P_{\text{gate}}$ of the capture rates $\gamma_1$ and $\gamma_2$, given by Eq.~(\ref{gamma12}). When $P_{\text{gate}}<3$~{\nano\watt}, i.e. $\gamma_1/\gamma_2<1$, an excess of holes and no electron populate the QD, and once $P_{\text{gate}}$ increases, the ratio $\gamma_1/\gamma_2$ increases and the exciton RE signal increases. When $P_{\text{gate}}=3$~{\nano\watt}, i.e. $\gamma_1/\gamma_2=1$, the QD is completely neutralized due to the symmetrical electron and hole capture rates, and the RE signal is maximum. It should be noted that the dynamical aspect of the process should not to be forgotten. When $P_{\text{gate}}>3$~{\nano\watt}, i.e. $\gamma_1/\gamma_2>1$, an excess of electrons and no hole populate the QD, and once $P_{\text{gate}}$ increases, the ratio $\gamma_1/\gamma_2$ increases and the exciton RE signal vanishes. The RE of the negative trion should then be observed. However, the bandwidth of the microcavity is not wide enough to detect any signal from this latter excitonic complex which is characterized by a typical binding energy of the order of \unit{7}{\milli\electronvolt} for InAs/GaAs QDs emitting at \unit{1.27}{\electronvolt} \cite{Schliwa:09}.

To conclude this section, the experimental study of the exciton RE over seven orders of magnitude of the gate power led to the determination of the nature of the carrier capture processes involved in a QD. While other groups \cite{Gomis:11,Ferreira:99} have previously used the non-resonant photoluminescence to evaluate the carriers capture rates, we benefit here from the resonant excitation and its optical gating effect to probe the QD ground state, which is determined by the capture of the electrons and holes created by the non-resonant laser. In particular, this method gives access to ultra-slow capture processes in the very low gate power regime where the non-resonant photoluminescence is completely negligible. Inversely, when the non-resonant photoluminescence is observed, the capture rates are already at least of the order of \unit{1}{\nano\reciprocal\second} \cite{Gomis:11}. In the context of our unusual experiments, the  population model that we used allows us to estimate the capture rates $\gamma_1$ and $\gamma_2$ which vary from a few {\milli\reciprocal\second} to a few hundreds of {\pico\reciprocal\second} in the explored power range (i.e. seven orders of magnitude). Finally, it should be noted that the estimations of $\gamma_1$ and $\gamma_2$ are in very good agreement with the theoretical calculations of the capture processes assisted by Auger effect and optical phonons emission. This study allows a first comprehensive confrontation between theory and experiment concerning the carriers capture in QDs.

\subsection{\label{theory_shift}Energy shifts of the neutral exciton and the positive trion}

The electric field created by a hole trapped in the vicinity of the QD induces, through the quantum confined Stark effect, energy shifts $\Delta_X$ on $E_X$ and $\Delta_{X^+}$ on $E_{X^+}$. The Stark effect is here induced by Coulomb interactions between the hole in the defect and the carriers in the QD, and $\Delta_X$ and $\Delta_{X^+}$ are obviously very sensitive to the relative position of the defect to the QD \cite{Jankovic:04,Abbarchi:08}. The energy shift of the $X$ and $X^+$ RE further depends on the number of holes in the defect and are simply given by:
\begin{equation}\label{energy X and X+}
    \delta E_X=n_p^{(st)}\Delta_X \text{~~and~~} \delta E_{X^+}=n_p^{(st)}\Delta_{X^+}
\end{equation}
where $n_p^{(st)}=2-\frac{2}{1+\frac{R_C+\gamma_3}{R_E+\gamma_4}}$ is the number of holes in the defect in the steady state regime, given by the rate equation~(\ref{trou_piege}).

The experimental results in figure~\ref{Figure3} are fitted with the expressions of $E_X$ and $E_{X^+}$ of Eq.~(\ref{energy X and X+}) by adjusting the following parameters:
\begin{eqnarray}
\left\lbrace
    \begin{array}{l}
        \Delta_X=7~{\micro\electronvolt}~\text{and}~\Delta_{X^+}=5~{\micro\electronvolt}\\
        R_C=1/30~{\micro\reciprocal\second}~\text{and}~R_E=1/60~{\micro\reciprocal\second}\\
        \gamma_{3({\micro\reciprocal\second})}=1/3P_{\text{gate}({\nano\watt})}\\
        \gamma_{4({\micro\reciprocal\second})}=0.3\sqrt{P_{\text{gate}({\nano\watt})}}\label{gamma34}
    \end{array}
\right.
\end{eqnarray}
Thoroughly, the values of $R_C$ and $R_E$ are deduced from the study of the charges dynamics in the transient regime when the optical gate is switched off, presented in section \ref{on to off}. Consequently, knowing these values allows us to determine the power dependence of $\gamma_3$ and $\gamma_4$. We deduce that the defect is never empty, even in the absence of the optical gate for which its steady state is characterized by $n_p^{(st)}|_{min}=n_p^{(st)}|_{\gamma_3=\gamma_4=0}\approx1$. The modification of the number of holes mainly happens in the very low optical gate power regime (i.e. $P_{\text{gate}}<10$~{\nano\watt}). Here, the red shift corresponds to a decrease of the number of holes in the defect whereas the blue shift is related to an increase of holes. The estimated values of $\Delta_X$ and $\Delta_{X^+}$ are used to evaluate the distance between the defect and the QD. Referring to the theoretical calculations of A. Jankovic \emph{et al.} of the Stark effect induced by a hole in the wetting layer \cite{Jankovic:04}, these shifts correspond to a distance of 20 to \unit{25}{\nano\meter} between the interface defect and the QD, which is also in good agreement with the work of M. Abbarchi \emph{et al.} \cite{Abbarchi:08}. This distance is large enough to avoid any overlap between the wave functions of the holes in the defect and in the QD. In other words, the holes are localized in the defect, resulting in small capture and escape rates $R_C$ and $R_E$.

We also explain that once the optical gate is turned on, the charge states of the defect and of the QD are no longer coupled since the tunneling channel becomes negligible (see section~\ref{pop with gate}). Therefore, the variations of the energies $E_X$ and $E_{X^+}$ are not correlated with the intensity variations of the $X$ and $X^+$ RE signals.

Finally, we discuss the power dependencies of the rates $\gamma_3$ and $\gamma_4$. For the trapping rate of holes in the defect $\gamma_3$, since the confinement energy of an interface defect $E_{\Phi}$ is very low, the hole capture by Auger effect is very efficient. The trapping rate $\gamma_3$ depends then essentially on Auger processes which are 60 times more efficient than the holes capture in the QD. Concerning the de-trapping rate $\gamma_4$ of holes from the defect to the continuum reservoir, either phonons absorption, or Auger processes involving one incident charge in the barrier and one target charge in the defect \cite{Bockelmann:92,Ferreira:99,Favero:07} which explain the square root power dependence, can be considered. On the one hand, the phonon-assisted mechanism would concern the absorption of acoustic phonons because the optical phonon states are not populated at low temperature \cite{Favero:07}. Moreover, contrarily to the case of charges trapped in the QD, the de-trapping from the defect by absorption of an acoustic phonon can be efficient due to the smaller confinement of the order of \unit{10}{\milli\electronvolt} which is comparable to the typical energy of the acoustic phonons involved with confined states. On the other hand, various experimental studies showed that the fluctuating electrostatic environment, related to the trapping and de-trapping of charges in defects close to a QD, is notably responsible for the zero-phonon line broadening in QDs at low temperature \cite{Berthelot:06,Favero:07}. In such context, the de-trapping of charges from the defects is driven by Auger effects, which are also very efficient for trapping holes in defects. We thus assume that $\gamma_4$ is, as $\gamma_3$, preferentially associated to Auger-assisted de-trapping processes, where one hole in the defect interacts with another hole in the barrier, than to the absorption of acoustic phonons.

\subsection{Study of the charges dynamics in the transient regime}

In the previous sections, we have studied the influence of the optical gate power on the intensity and the energy of the RE signal, which led to analyze, in the steady state regime, the charge state of the QD and of the interface defect in its vicinity, respectively. In the corresponding experiments, the optical gate is always switched on and the physical properties are average values in the steady state regime. In this section, we focus on the dynamics of the optical gate effect by studying the transient regimes where the optical gate has just been switched on or off. We consider two situations: the OFF stage when the optical gate is turned off at $t=0$, knowing that the system was in a steady state with the applied gate at $t<0$; and the ON stage when the optical gate is switched on at $t=0$, knowing that the system without applied gate was in a steady state at $t<0$.

Experimentally, we send two light beams. The first one, provided by the tunable cw external cavity laser diode, resonantly excites the excitonic transition, while the second one is our optical gate (HeNe laser), which is modulated from 0 to few nWs at \unit{400}{\hertz} by using an acousto-optic modulator. An oscilloscope records a time-histogram of the RE where each detected photon corresponds to one event. The temporal resolution of the experimental setup is evaluated by measuring the system response for the two stages when the modulated optical gate is directly sent to the photo-detector, and turns to be \unit{1.7}{\micro\second} in the OFF stage and \unit{2}{\micro\second} in the ON stage.

\subsubsection{\label{on to off}OFF stage}

Under resonant excitation, two effects are responsible for the RE intensity variation when the optical gate is switched off at time $t=0$: the energy shift of the transition which depends on the gate power as described in sections~\ref{shift} and \ref{theory_shift}; and the quenching of the exciton RE due to the charging of the QD by holes tunneled from the defect to the QD at a rate $\gamma_{in}$. These two phenomenons are governed by the time constants $t_{\text{shift}}$ and $t_{\text{off}}$ which affect the central energy of the transition and the recorded RE intensity, respectively. The physical meaning of these two times appears clearly when considering the RE spectrum, which can be recorded. For $t>0$ and at a given detuning $\delta$ between the resonant excitation laser and the exciton energies, it can be modeled by a Lorentzian line:
\begin{equation}
    I(\delta,t>0)=\frac{I_0\text{e}^{-t/t_{\text{off}}}}{[\delta-\Delta(1-\text{e}^{-t/t_{\text{shift}}})]^2+\Gamma'^2/4}
\end{equation}

\begin{figure}[htb]
\begin{center}
\includegraphics[width=8.6cm]{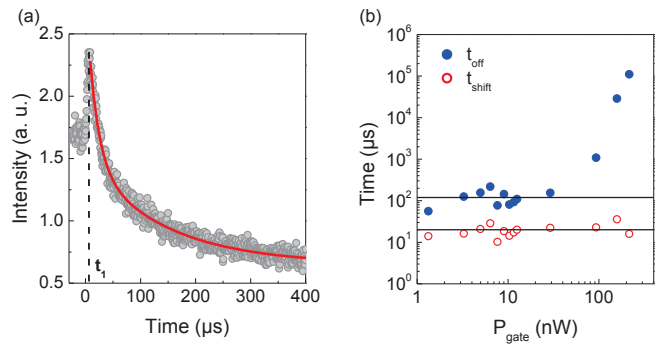}
\end{center}
\caption{\label{Figure8}{(a) Temporal evolution of the neutral exciton resonant emission when the optical gate is switched off at $t=0$ ($P_{\text{gate}}=3$~{\nano\watt}). The detuning between the resonant excitation laser and the exciton energy is equal to $\delta=-1$~{\micro\electronvolt}. (b) Time constants $t_{\text{shift}}$ and $t_{\text{off}}$ as a function of the optical gate power $P_{\text{gate}}$.}}
\end{figure}

One example of the temporal evolution of the neutral exciton RE is shown in figure~\ref{Figure8}(a), for an optical gate power of \unit{3}{\nano\watt}, at $\delta=-1$~{\micro\electronvolt}. Two distinct features are observed in the experimental results. For $0<t\leq t_1$, the exciton energy $E_X$ decreases and it thus gets closer to the resonant laser energy $E_{L}$ because of the initially chosen negative detuning. Therefore, the RE signal starts to increase. At $t=t_1$, the laser is strictly resonant with the excitonic transition and the RE signal is maximum. When $t>t_1$, the red shift phenomenon is still effective and the exciton energy $E_X$ moves away from the resonant excitation laser energy $E_{L}$. The RE signal then first decreases with the time constant $t_{\text{shift}}$. Once the exciton energy reaches its limit ($t\gg t_{\text{shift}}$), the RE signal still slowly decreases with the time constant $t_{\text{off}}\gg t_{\text{shift}}$, really characteristic of the dynamics of the RE intensity. For $\delta=-1$~{\micro\electronvolt}, the experimental data are fitted (Fig.~\ref{Figure8}(a), solid line) by a double exponential decay with the time constants $t_{\text{shift}}=20\pm1$~{\micro\second} and $t_{\text{off}}=150\pm10$~{\micro\second}. Performing the same experiment for various detunings and optical gate powers allows a complete study of the variations of $t_{\text{off}}$ and $t_{\text{shift}}$ with the gate power, as well as a direct
comparison with our theoretical model. The results that are presented in figure~\ref{Figure8}(b) first show that $t_{\text{shift}}$ is relatively constant over the whole explored range of gate power ($P_{\text{gate}}\leq200$~{\nano\watt}). For $t_{\text{off}}$, we observe also a relative constant value for $P_{\text{gate}}\leq30$~{\nano\watt} followed by a strong increase once $P_{\text{gate}}\geq100$~{\nano\watt}, until a remarkable value of \unit{20}{\milli\second} for $P_{\text{gate}}\sim200$~{\nano\watt}.

When the optical gate is switched off, the remaining photo-generated carriers in the barrier can relax through two channels: a direct radiative recombination of the electron-hole pairs with a radiative lifetime of about \unit{1}{\nano\second} \cite{Gobel:85}; and a non-radiative relaxation towards the QDs or the defects with time constants of the order of \unit{10}{\micro\second} (given by the capture rates values $\gamma_1$, $\gamma_2$, $\gamma_3$ at $P_{\text{gate}}=3$~{\nano\watt}, and $R_C$). Since for ultra-low carrier populations, such as the ones we consider now, the second channel is much less efficient than the first one, we assume that, for an initial very low gate power, after switching off the optical gate, the barrier empties almost instantaneously by radiative recombination of the electron-hole pairs. In other words, neither a modification of the charge state of the QD nor of the defect is induced by the GaAs barrier. The charge state of the QD and of the defect can then be described as follows.

The hole population in the defect at anytime $t$ is deduced from the rate equation~(\ref{trou_piege_gate_off}), where $\gamma_{in}$ and $\gamma_{out}$ are neglected with respect to the trapping and de-trapping rates $R_C$ and $R_E$, and is given by
\begin{equation}
    n_p(t)\approx\left(n_p(0)-\frac{2R_C}{R_C+R_E}\right)\text{e}^{-(R_C+R_E)t}+\frac{2R_C}{R_C+R_E}
\end{equation}
When the optical gate is switched off, the defect relaxes to its steady state defined by $n^{(st)}_p=2R_C/(R_C+R_E)$. Since the neutral exciton energy is defined as $\delta E_X(t)=n_p(t)\Delta_X$ (see Eq.~(\ref{energy X and X+})), $E_X$ is time dependent and follows an exponential law with a time constant $t_{\text{shift}}=(R_C+R_E)^{-1}$. This energy shift can be either a red shift or a blue shift depending on the initial gate power.

As far as the QD population is concerned, when the optical gate is turned off, the hole transfer to the QD is ensured by the tunneling channel between the QD and the defect which is not negligible anymore since there is no more charges in the barrier. Moreover, since $\gamma_{out}\ll\gamma_{in}\ll R_C,R_E$, the hole capture in the QD is much slower than the recovery of the defect's equilibrium. Therefore, we can assume that the number of holes in the defect has already reached its steady state value $n_p^{(st)}$ when the number of holes in the QD starts to increase. The rate equation~(\ref{trou_boite_gate_off}) then leads to the following hole population in the QD at anytime $t$,
\begin{equation}
    n_h(t)\approx2-(2-n_h(0))\text{e}^{-\gamma_{in}n_p^{(st)}t}
\end{equation}
where the hole transfer to the QD is done at a rate $\gamma_{in}n_p^{(st)}$ which defines the time constant $t_{\text{off}}=\left(\gamma_{in}n_p^{(st)}\right)^{-1}$. When the optical gate is turned off, the QD hole population relaxes to its steady state value defined by $n_h^{(st)}\approx2$.

To summarize, in the context of an initial ultra-weak optical gate, the time constants $t_{\text{shift}}$ and $t_{\text{off}}$ are given by:
\begin{equation}
    t_{\text{shift}}=\frac{1}{R_C+R_E}\text{~~and~~}t_{\text{off}}=\frac{1}{\gamma_{in}}\frac{R_C+R_E}{2R_C}
\end{equation}
Let us first consider the large increase of $t_{\text{off}}$ for initial gate powers larger than $100$~{\nano\watt}. The initial carrier population in the barrier starts to be sufficiently large so that our main hypothesis of an instantaneously emptied barrier through radiative recombination breaks down. An increasing amount of carriers relax towards the QD and the defect, substantially altering the charging dynamics of the QD. We emphasize that the direct relaxation of electrons towards the QD will essentially slow down the retrieval of the equilibrium situation. For gate powers smaller than $100$~{\nano\watt}, the measurement of the time constant $t_{\text{shift}}\approx20$~{\micro\second} (average value as a solid line in figure~\ref{Figure8}(b)) allows to estimate the values of $R_C=1/30$~{\micro\reciprocal\second} and $R_E=1/60$~{\micro\reciprocal\second} which were presented previously in section~\ref{theory_shift}.

Let us now come back to the hierarchy introduced for the relaxation rates in section~\ref{pop with gate} (Eq.~(\ref{hierarchy})). In the low optical gate power regime, the knowledge of the time constant $t_{\text{off}}\approx120$~{\micro\second} (average value as a solid line in figure~\ref{Figure8}(b)) allows to give an estimate of the capture rate from the defect to the QD, which turns to be  $\gamma_{in}\approx1/160$~{\micro\reciprocal\second} (and $\gamma_{in}\gg\gamma_{out}$ at low temperature). Considering the estimate of $R_C=1/30$~{\micro\reciprocal\second}, $R_E=1/60$~{\micro\reciprocal\second}, and $\gamma_{1}\approx 1/25$~{\micro\reciprocal\second}, $\gamma_{2}\approx 1/15$~{\micro\reciprocal\second}, $\gamma_{3}\approx 1/3$~{\micro\reciprocal\second} and $\gamma_{4}\approx 1/3$~{\micro\reciprocal\second} for a typical small gate power $P_{\text{gate}}\approx1$~{\nano\watt}, the equation~(\ref{hierarchy}) is verified. Moreover, since all these rates (except $R_C$ and $R_E$ which are constant) increase with gate power, the hierarchy introduced previously is fulfilled for a large range of gate powers, at least corresponding to our experimental conditions.

\subsubsection{ON stage}

\begin{figure}[htb]
\begin{center}
\includegraphics[width=8.6cm]{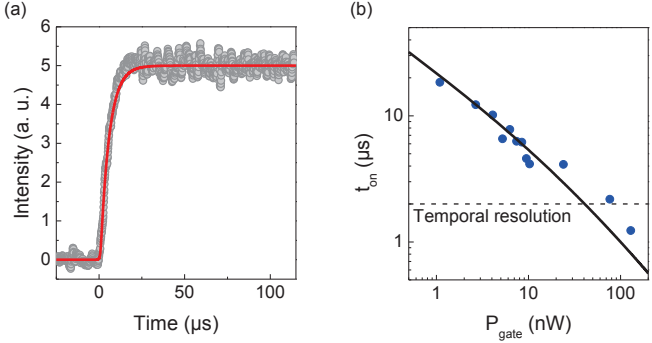}
\end{center}
\caption{\label{Figure9}{(a) Temporal evolution of the neutral exciton resonant emission when the optical gate is switched on at $t=0$ ($P_{\text{gate}}=10$~{\nano\watt}). The detuning between the resonant excitation laser and the exciton energy is equal to $\delta=-1$~{\micro\electronvolt}. (b) Time constant $t_{\text{on}}$ as a function of the optical gate power $P_{\text{gate}}$. The theoretical fit (solid line) is given by the power laws~(\ref{gamma12}) of the capture rates $\gamma_1$ and $\gamma_2$.}}
\end{figure}

We now present the same set of measurements for the ON stage. Figure~\ref{Figure9}(a) shows an example of the temporal evolution of the neutral exciton RE when the optical gate is switched on ($P_{\text{gate}}=10$~{\nano\watt}), at the same detuning $\delta=-1$~{\micro\electronvolt} chosen for the OFF stage experiment. For $t<0$, the detected signal is negligible since the optical gate is switched off, whereas for $t>0$, the RE signal increases very quickly as soon as the optical gate is switched on. Contrarily to the OFF stage, no energy shift of the exciton is observed. Indeed, as described in section~\ref{theory} (see Fig.~\ref{Figure5}), when the optical gate is switched on, the charge state in the defect is mainly governed by the trapping and de-trapping processes (3) and (4) characterized by the time constants $\gamma_3^{-1}\sim0.1$~{\micro\second} and $\gamma_4^{-1}\sim1$~{\micro\second} at $P_{\text{gate}}=10$~{\nano\watt} (Eq.~(\ref{gamma34})). With such timescales, as the time resolution of our gate modulation experiment is limited to \unit{2}{\micro\second}, the intensity rise induced by the energy shift is beyond our detection limit. The experimental data in figure~\ref{Figure9}(a) are well fitted by the convolution of one exponential with a time constant $t_{\text{on}}=5\pm1$~{\micro\second} and the system time response of \unit{2}{\micro\second}.

We now consider the dependence of $t_{\text{on}}$ on the optical gate power, shown in figure~\ref{Figure9}(b). The time constant $t_{\text{on}}$ decreases with increasing $P_{\text{gate}}$. Indeed, the appearance of the RE signal once the optical gate is turned on is related to the capture processes (1) and (2) characterized by the rates $\gamma_1$ and $\gamma_2$ which vary with the optical gate power (see section~\ref{theory RE vs gate power}). We then expect that the time constant $t_{\text{on}}$ strongly depends on $P_{\text{gate}}$. Thanks to the resolution of the linear differential system~(\ref{mpa}) developed within the population evolution model and the expressions~(\ref{gamma12}) of $\gamma_1$ and $\gamma_2$, we deduce that $t_{\text{on}}=\frac{3}{2(\gamma_1+\gamma_2)}$. We see in figure~\ref{Figure9}(b) that this latter expression fits well our experimental data.

\subsection{\label{g2}Photon auto-correlation}

\subsubsection{Experimental results}

\begin{figure}[htb]
\begin{center}
\includegraphics[width=8.6cm]{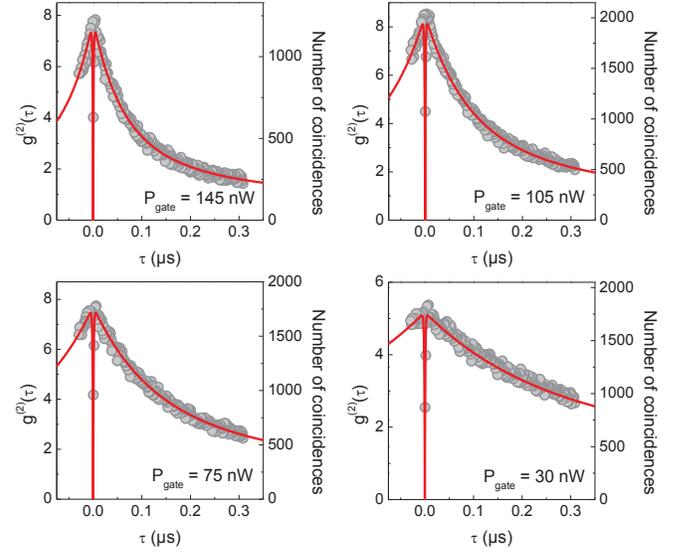}
\end{center}
\caption{\label{Figure10}{Intensity auto-correlation measurements of the optically-gated resonant emission of the neutral exciton, for 4 optical gate powers (taken from a complete study over 12 gate powers). The experimental data correspond to the number of coincidences as a function of the time delay while the theoretical curves, results of the population evolution model, are plotted on a dual scale by adjusting the experimental and theoretical maxima. The parameters used in the theoretical model are $\gamma=3030$~{\micro\reciprocal\second}, $r_X=500$~{\micro\reciprocal\second}. $\gamma_1$ and $\gamma_2$ are given by the power laws~(\ref{gamma12}).}}
\end{figure}

In order to study the photon statistics of the optically-gated RE of the neutral exciton, the intensity auto-correlation function of the emitted photons is measured by using  a conventional Hanbury-Brown and Twiss setup. This experiment is done when the QD is in its steady state (i.e. the optical gate is always turned on) for various optical gate powers in the range \unit{20-200}{\nano\watt}. The results of 4 measurements (over a set of 12 measurements) are presented in figure~\ref{Figure10}. Even if the time window has a width of \unit{350}{\nano\second}, which is large for photon correlation experiments in InAs/GaAs QDs \cite{Michler:00,Zwiller:01,Yuan:02}, it appears not large enough to observe the limit value of the number of coincidences at long time scales, and therefore, the data normalization becomes tricky. To overcome this problem, the raw experimental data are plotted in the unit of the number of coincidences as a function of the time delay $\tau$. Another consequence of this large time window is the poor resolution of  the experimental antibunching dip observed at zero delay.

The antibunching at zero delay is the signature that the RE signal corresponds to the single photon emission from the QD \cite{Nguyen:11}. Apart from zero delay, on all spectra, a strong bunching appears, up to values of $g^{(2)}(\tau)=8$ for $P_{\text{gate}}=145$~{\nano\watt}. We note that this bunching decays with a time constant  $\tau_B\sim0.1$~{\micro\second} (for $P_{\text{gate}}=145$~{\nano\watt}), much longer than the radiative lifetime $T_1=330$~{\pico\second} of the QD. In fact, the $g^{(2)}(\tau)$ function reflects the population dynamics of the neutral exciton state which is given, in our model, by the probability ${\cal P}_{1_h;1_e}$. This dynamics is driven not only by the radiative recombination but also by the capture processes occurring on very long times $\gamma_1^{-1}$ and $\gamma_2^{-1}$. As shown in the following, the capture rates of the carriers photo-generated in the barrier by the optical gate are responsible for the bunching phenomenon. Therefore, the order of magnitude of the measured time constant $\tau_B$ is given by $\gamma_1^{-1}$ and $\gamma_2^{-1}$, and will give values of $\tau_B$ up to few hundreds of {\nano\second}.

\subsubsection{Photon statistics of the resonant emission}

\begin{figure}[htb]
\begin{center}
\includegraphics[width=8.6cm]{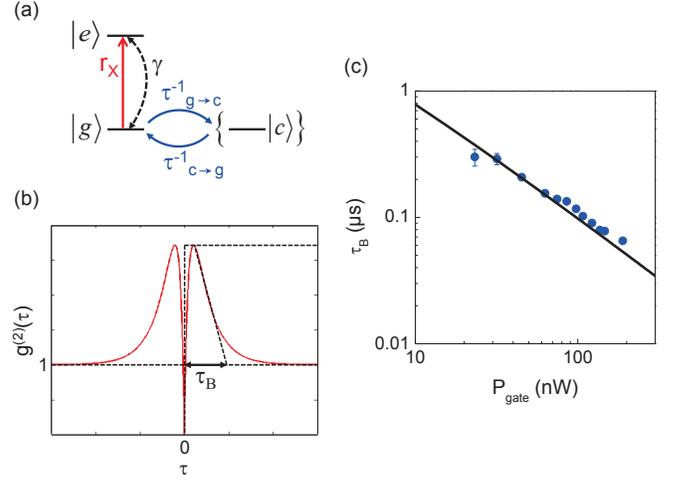}
\end{center}
\caption{\label{Figure11}{(a) Sketch showing the fluctuation of the quantum dot charge state and its resonant emission when the resonant laser and the optical gate are simultaneously switched on. (b) Trend of the second order correlation function of the exciton resonant emission characterized by the bunching time constant $\tau_B$. (c) Experimental values of the time constant $\tau_B$ characterizing the bunching decay as a function of the optical gate power $P_{\text{gate}}$ (symbols) and theoretical curve corresponding to the equation~(\ref{bunching time}) (solid line).}}
\end{figure}

In the regime of very low optical gate power, where $\gamma_1,\gamma_2\ll\gamma$, the emitted photons come mostly from the RE because the non-resonant photoluminescence excited by the optical gate is negligible. However, the resonant laser only excites the QD if this latter is in its empty ground state $|0_h;0_e\rangle$. Therefore, the photon statistics strongly depends on the QD charge state. As discussed in the theoretical model, the QD charge state fluctuates over time due to the charge capture processes (1) and (2) characterized by the rates $\gamma_1$ and $\gamma_2$. This fluctuation is completely described by our model with its 9 possible charge states (see Fig.~\ref{Figure6}). As depicted in figure~\ref{Figure11}(a), once the QD is empty, the QD remains uncharged for a time $\tau_{g\rightarrow c}$. After a time $\tau_{g\rightarrow c}$, the QD evolves towards one of its charged states $\{|c\rangle\}$. The charged state $|c\rangle$ presents either an excess of electrons with no hole if $\gamma_1>\gamma_2$ or an excess of holes with no electron if $\gamma_1<\gamma_2$. From the scheme of Fig.~\ref{Figure6}), we obtain, in first approximation, that:
\begin{eqnarray}\label{tau_up}
\tau_{g\rightarrow c}^{-1}\approx
\left\lbrace
    \begin{array}{l}
        2\gamma_1\text{~~if~~}\gamma_2<\gamma_1\ll\gamma\\
        2\gamma_2\text{~~if~~}\gamma_1<\gamma_2\ll\gamma
    \end{array}
\right.
\end{eqnarray}
Similarly, the QD remains in a charged state $\{|c\rangle\}$ during a time $\tau_{c\rightarrow g}$ (the QD can evolve from one charged state to another). Since, in the regime of very low gate power, the disappearance of a hole (an electron) arises from the slow capture of an electron (a hole) followed by their fast radiative recombination, we deduce from figure~\ref{Figure6} that the time constant $\tau_{c\rightarrow g}$ is approximately given by:
\begin{eqnarray}\label{tau_down}
\tau_{c\rightarrow g}^{-1}\approx
\left\lbrace
    \begin{array}{l}
        2\gamma_2\text{~~if~~}\gamma_2<\gamma_1\ll\gamma\\
        2\gamma_1\text{~~if~~}\gamma_1<\gamma_2\ll\gamma
    \end{array}
\right.
\end{eqnarray}

Let us now discuss the way  the QD charge state is probed by the resonant laser at a pumping rate $r_X$. Under resonant excitation, as soon as the QD is in its ground state $|g\rangle$, it is resonantly pumped to its excited state $|e\rangle$ (see Fig.~\ref{Figure11}(a)). Once in the excited state, the QD emits a photon and ends up in its ground state. This cycle is repeated during the time $\tau_{g\rightarrow c}$ after which the QD evolves to a charged state $\{|c\rangle\}$, and the resonant laser can no longer excite the QD at the neutral exciton energy. The single photon emission is then blocked for the time $\tau_{c\rightarrow g}$ while the QD remains in a charged state. After $\tau_{c\rightarrow g}$, the QD is back to its ground state and the resonant laser plays its role again. Therefore, the QD emits packets of single photons with a statistics governed by the exchange dynamics between the ground state $|g\rangle$ and the charged states $\{|c\rangle\}$. We note that the time intervals characterizing the emission of the photons packets are very long compared to the QD radiative lifetime since $\gamma\gg\gamma_1,\gamma_2$.

The probability of detecting two photons for a delay $\tau$ of the order of the bunching time $\tau_B$ is high. It is possible to show that $\frac{1}{\tau_B}=\frac{1}{\tau_{g\rightarrow c}}+\frac{1}{\tau_{c\rightarrow g}}$ \cite{Sallen:10,Sallen:11} and thus, in our case:
\begin{equation}
    \tau_B=\frac{1}{2(\gamma_1+\gamma_2)}\label{bunching time}
\end{equation}
The bunching time $\tau_B$ is obtained from an exponential fit of the  experimental results for $g^{(2)}(\tau)$, as schematized in figure~\ref{Figure11}(b). Figure~\ref{Figure11}(c) shows the values obtained for $\tau_B$ as a function of the optical gate power. These values are in fair agreement with the equation~(\ref{bunching time}), where $\gamma_1$ and $\gamma_2$ are given by the power laws (\ref{gamma12}) presented in section~\ref{theory RE vs gate power}. This confirms the validity of our approach.

Moreover, the entire set of measurements of the intensity auto-correlation function can be fitted in the framework of our model. For the considered two-level system, where the excited state is the excitonic state $|1_h;1_e\rangle$ and the ground state is the vacuum state $|0_h;0_e\rangle$, the probability of detecting a single photon at delay $\tau$ is proportional to the population of the excited state at delay $\tau$ or in other words to the probability ${\cal P}_{1_h;1_e}(\tau)$. Likewise, the probability of detecting a single photon at zero delay is proportional to the probability of having the system in its ground state and consequently to the probability ${\cal P}_{0_h;0_e}(\tau=0)=1$, the others being set to zero. The $g^{(2)}(\tau)$ function can then be defined as the probability of detecting a photon at delay $\tau$, knowing that a photon has been detected at zero delay, normalized to the total probability ${\cal P}^{(st)}_{1_h;1_e}$ of detecting a photon due to the $X$ recombination:
\begin{equation}
    g^{(2)}(\tau)=\frac{{\cal P}_{1_h;1_e}(\tau)}{{\cal P}^{(st)}_{1_h;1_e}}\Bigg\lvert_{{\cal P}_{0_h;0_e}(\tau=0)=1}
\end{equation}
The theoretical curves of $g^{(2)}(\tau)$, calculated with the previously determined  $\gamma_1$ and $\gamma_2$ values, are displayed in figure~\ref{Figure10}. As already stated, our time window is not large enough to observe the limit value of the number of coincidences at long time scales and in figure~\ref{Figure10} the number of recorded coincidences is plotted as a function of the delay. The maxima of the experimental and theoretical curves are then displayed on a double scale. Once again we observe a good agreement between the experimental findings and the predictions of our model.

The measurements of the second order auto-correlation function allow to establish a clear link between the observed bunching effect and the capture processes in the QD of the carriers that are photo-generated by the optical gate. Under resonant excitation, in the very low gate power regime, the single photon emission occurs during a characteristic time given by $\tau_{g\rightarrow c}\approx\frac{1}{\text{max}(2\gamma_1,2\gamma_2)}$ (Eq.(\ref{tau_up})) and the time interval between two consecutive single photons packets is given by $\tau_{c\rightarrow g}\approx\frac{1}{\text{min}(2\gamma_1,2\gamma_2)}$ (Eq.(\ref{tau_down})). This induces a blinking effect of the RE. Such a behavior may be reminiscent of the one observed on the non-resonant photoluminescence in QDs and nanocrystals. Although slow bunching effects were observed, they arose from the so-called spectral diffusion effect \cite{Berthelot:06,Sallen:10,Robinson:00,Besombes:02,Abbarchi:12}. In our case, the QD charge state fluctuations is the main reason for photon bunching \cite{Santori:04,Pietka:13} and we observe a blinking between distinct excitonic states, namely the neutral exciton and the other excitonic complexes. Note that a similar blinking behavior had already been observed in the differential transmission of a single QD under resonant excitation \cite{Hogele:04}.

\section{CONCLUSION}

To summarize, we have shown that the electrostatic environment plays a crucial role on the optical response of a single QD even under resonant excitation where no charge is photo-created in its vicinity. A self-consistent calculation presented in appendix highlighted the influence of the unintentional residual carbon doping under no optical excitation which leads to the presence of at least one residual charge in the QD. As a result, more than 90\% of the QDs present a complete quenching of the neutral exciton RE because of Coulomb blockade effect. In this context, an additional weak non-resonant laser is used as an optical gate to control the QD charge state thanks to the capture of the photo-created charges in the GaAs barrier, resulting in the carrier draining of the QD. The optical gate not only induces a complete retrieval of the neutral exciton RE, but also allows an efficient control of the RE in the original regime where the photoluminescence non-resonantly excited by the optical gate is completely negligible. We developed a population evolution model where the different capture and escape processes induced by the optical gate are considered to describe the dynamics of the QD charge state. This model perfectly fits, with only a single set of reasonable parameters, the various experimental results presented over seven orders of magnitude of the optical gate power, in the steady state and dynamical regimes. It appears that the QD ground state, and thus the RE intensity, is governed by peculiar Auger- and optical phonon-assisted capture processes characterized by long relaxation times of the order of 1 to \unit{100}{\micro\second}. Moreover, the measurements of the capture coefficients for both processes are in good agreement with theoretical calculations of various research groups. This study then constitutes a first confrontation between theory and experiment with such a low dynamic, concerning the carriers capture in QDs. Finally, the slow capture processes can be directly associated to an intermittent behavior of the QD RE, which is similar to the blinking effect in QDs and nanocrystals, and where the RE quenching will be observed as long as the QD ground state is a charged state. In summary, we presented an experimental and theoretical study where the optical gate acts as a very sensitive probe of the residual doping and the QD population in an unprecedented explored regime of weak non-resonant excitation.

\begin{acknowledgements}
The authors gratefully acknowledge P. M. Petroff for providing the sample, and G. Bastard, C. Delalande and L. Lanco for fruitful discussions and comments. The work was supported by the ANR (Agence Nationale de la Recherche) project EXTREME.
\end{acknowledgements}

\appendix*\section{Influence of the residual doping on the charge state of the quantum dot}

\begin{figure}[htb]
\begin{center}
\includegraphics[width=8.6cm]{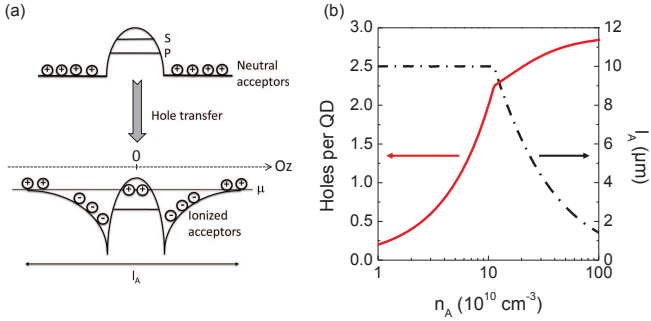}
\end{center}
\caption{\label{FigureAppendix}{(a) Top: Sketch along the growth direction $(0z)$ of the valence band of an empty QD surrounded by neutral acceptors from the unintentional residual carbon doping. Bottom: Sketch of the same QD after the acceptors ionization and the resulting holes transfer towards the QD. $l_A$ is the thickness of depletion region around the QD plane and $\mu$ the chemical potential fixed to the Fermi energy of the neutral acceptors outside the depletion region. (b) Hole occupancy and depletion length $l_A$ as a function of the bulk doping density $n_A$, for a QD density $n_{\text{QD}}=5\times10^{7}$~{\centi\meter\rpsquared}, at $T=7$~{\kelvin}.}}
\end{figure}

According to the experiments, the sample has an unintentional residual carbon doping (C-doping) \cite{Hayne:04} of volume density $n_A$. The carbon binding energy $E_A$ is about \unit{26.7}{\milli\electronvolt}, so that thermal activation can be neglected at the temperature of the experiments (\unit{7}{\kelvin}). As a consequence, in absence of the QDs, the acceptors are neutral and the Fermi energy is pinned at the energy $E_{A}$ above the top of the GaAs valence band. A QD introduces bound states well above this energy, and thus some holes transfer towards the QDs, leading to the formation of a depletion region around the QD plane (see Fig.~\ref{FigureAppendix}(a)). To estimate the density of transferred holes, we assume an idealized ensemble of identical QDs (of areal density $n_{\text{QD}}$ around $10^{8}$~{\centi\meter\rpsquared}) uniformly distributed in the QD plane (taken at $z=0$). Neglecting both the QD height along the growth axis ($(0z)$ direction) and the structuration due to the Bragg mirrors, one obtains a volume charge distribution in the sample:
\begin{equation}\label{charge density}
\rho(z)=\left\lbrace
    \begin{array}{cc}
        -en_A +ep_B\delta(z)&\text{for~} |z|<l_{A}/2<l_M/2\\
        0&\text{otherwise}
    \end{array}
\right.
\end{equation}
where $l_{A}$ is the thickness of the depletion layer around the QD plane (Fig.~\ref{FigureAppendix}(b)) and $p_{B}$ the areal density of bound (to the QDs) holes. We assume that acceptors are present only in a finite region around the QD plane, of total thickness $l_M$ (for definiteness, we consider the thickness of the epitaxially grown cavity and take $l_M\approx10$~{\micro\meter} in the calculations). Charge neutrality gives $n_{A}l_{A}=p_{B}$ if only part of acceptors are ionized ($l_A<l_M$), or $n_Al_M=p_B$ if all holes from doping are transferred into the QDs ($l_A=l_M$). Finally, the one-dimensional Poisson equation imposes a red shift of the QD levels by:
\begin{equation}\label{energy_shift}
    E_{SC}= -\frac{e^{2}p_{B}^{2}}{ 8\epsilon_{0}\epsilon_{r}n_{A}}
\end{equation}
where $\epsilon_{r}=12.5$ is the GaAs relative dielectric constant.

We consider that the QDs have cylindrical symmetry and, for the parameters in our sample, can bind up to 6 holes, distributed in the S and P shells (respectively twice and fourfold degenerated, considering both orbital and spin degrees of freedom). The S and P single electron and hole states are initially variationally calculated for a QD of in-plane radius $R=75$~{\angstrom} and height $h=15$~{\angstrom} (for which one calculates $E_X\approx1274$~{\milli\electronvolt}). The one-hole states are then used for the calculation of the coulombic-correlated (and fully anti-symmetrized) hole states of the multi-charged QD (hosting $N=2,...,6$ holes; for details see \cite{Isaia:03}). Finally, we evaluate the average hole occupancy per QD:
\begin{equation}\label{eta}
    \frac{p_B}{n_{\text{QD}}}=\frac{\sum\limits_{N}N\sum\limits_{n(N)}\exp\left[\beta\left(N\mu-E_{n(N)}\right)\right]}{\sum\limits_{N}\sum\limits_{n(N)}\exp\left[\beta\left(N\mu-E_{n(N)}\right)\right]}
\end{equation}
where $E_{n(N)}$ is the energy of the $n^{\text{th}}$ state of the QD with $N$ holes ($N=0, 1,...,6$), $1/\beta = k_{B}T$ and $\mu$ is the chemical potential. Self-consistency of the hole occupancy is obtained by rigidly red-shifting the one-hole energies entering in the calculations of $E_{n(N)}$ by $E_{SC}$ (Eq.~(\ref{energy_shift})). Finally, the value of $\mu$ is either imposed to be equal to the energy of holes bound to the acceptors outside the depletion region if the transfer is partial ($l_A<l_M$), or should be found is a self-consistent way when the transfer is maximum ($l_A=l_M$).

Figure~\ref{FigureAppendix}(b) shows the calculated hole occupancy (left scale) and depletion layer thickness (right scale) as a function of the bulk doping density $n_A$ (at $T=7$~{\kelvin}) and for a QD density of $5\times10^{7}$~{\centi\meter\rpsquared}. We clearly observe the consecutive filling of the various valence-band shells with increasing acceptor density. It is worth stressing that at low doping densities, all holes transfer to the QD plane and, because we focus on the low QD density region of the sample, the number of holes per QD is rather important: we obtain for instance that two holes dwell in the QDs for $n_{A}\approx10^{11}$~{\centi\meter\rpcubed}. For larger C-doping densities, the QD occupancy increases more slowly with increasing $n_A$, while at the same time the depletion length rapidly decreases below it maximum value. For the typical acceptor densities of the order of $10^{13}-10^{14}$~{\centi\meter\rpcubed} \cite{Moskalenko:03}, the QD is nearly populated by three holes, which is more than the two holes occupancy considered in our work hypothesis (see section~\ref{pop without gate}). However, excepting an average value of the typical acceptor densities, we do not have a precise knowledge of the C-doping in the QD vicinity. Note finally that notwithstanding for low QD and acceptor densities, the model described by Eqs.~(\ref{charge density}) to (\ref{eta}) is very crude (both dot-to-dot and inter-acceptor mean distances become very large), it nevertheless correctly describes the average occupancy $p_B=n_Al_M$ in the full transfer regime (i.e., when $n_Al_M/n_{\text{QD}}<1$), which could be anticipated from a more general standpoint (i.e., statistical distribution with charge neutrality consideration).

\end{document}